\documentclass{sf2a-conf2019}
\usepackage{graphicx}
\usepackage{hyperref}
\usepackage[]{natbib}  
\usepackage{epstopdf}
\usepackage[varg]{txfonts}
\usepackage{color}
\usepackage{rotating}

\def\BibTeX{{\rm B\kern-.05em{\sc i\kern-.025em b}\kern-.08em
    T\kern-.1667em\lower.7ex\hbox{E}\kern-.125emX}}
\bibpunct{(}{)}{;}{a}{}{,}  

\begin{document}

\TitreGlobal{SF2A 2021}


\title{The best broths are cooked in the oldest pans: revisiting the archival HST/FOC observations of quasars}

\runningtitle{Revisiting the HST/FOC observations of quasars}

\author{F. Marin}\address{Universit\'e de Strasbourg, CNRS, Observatoire astronomique de Strasbourg, UMR 7550, F-67000, Strasbourg, France}

\author{T. Barnouin$^1$}

\author{E. Lopez-Rodriguez}\address{Kavli Institute for Particle Astrophysics and Cosmology (KIPAC), Stanford University, USA}

\setcounter{page}{237}

\maketitle

\begin{abstract}
The Faint Object Camera (FOC) aboard the Hubble Space Telescope (HST) observed 26 individual active galactic nuclei (AGNs) in ultraviolet imaging polarimetry between 1990 and 2002.
Tremendous progresses have been made thanks to those high spatial resolution, high signal-to-noise ratio observations, such as the identification of the location of hidden active nuclei
and the three dimensional arrangement of polar material within the first hundred of parsecs around the central core. However, not all AGN observations have been reduced and analyzed, 
and none in a standardized framework. In this lecture note, we present our project of downloading, reducing and analyzing all the AGN HST/FOC observations that were achieved using
a consistent, novel and open-access reduction pipeline. We briefly present our methodology and show the first, preliminary result from our reduction pipeline: NGC~1068.
\end{abstract}

\begin{keywords}
Instrumentation: polarimeters, Methods: observational, Polarization, Astronomical data bases, Galaxies: active, Galaxies: Seyfert
\end{keywords}


\section{Introduction}
Ultraviolet (UV) observations allowed to make a huge leap forward in the field of astrophysics. UV pictures of the cosmos are rather different from the familiar collection of stars and 
galaxies seen at optical wavelengths. Main sequence stars become dimmer, since they emit most of their bolometric luminosity in the visible or near-infrared bands. Newly formed high-mass 
stars, producing UV radiation and violent stellar winds, light up instead. Similarly, irregular or spiral (young) galaxies and elliptical (old) galaxies become more visible because 
of the extra-production of UV light. Indeed, young galaxies experiment strong star formation, leading to large UV fluxes. In the case of old galaxies, UV photons are mainly produced by dying stars 
that have shed their cool outer layers in the post-red-giant phase of their evolution, revealing their small, hot cores \citep{Code1979,Davidsen1993}. It follows that UV astronomy can 
probe the evolution of stars and galaxies, in the early and late stages of their evolution. Those observations are essential to understand how the Universe has evolved from its origins to 
nowadays.

The Hubble Space Telescope (HST) was the first major space telescope to reveal the near, mid and far UV spectrum of the sky, though other UV instruments have flown on smaller observatories 
such as GALEX, as well as sounding rockets and the Space Shuttle \citep{Linsky2018}. Among the original instruments aboard the HST, the Faint Object Camera (FOC) was a particular device.
It consisted of a long-focal-ratio, photon-counting imager capable of taking high-resolution images in the wavelength range 1150 -- 6500~\AA. When corrected by COSTAR, the field-of-view
(FoV) and pixel size of the f/96 camera were 7" $\times$ 7" (512 $\times$ 512 format) and 0.014" $\times$ 0.014", respectively. But, most importantly, it was a polarimeter. The huge spatial
resolution offered by the FOC, coupled to the very low instrumental polarization and excellent polarizing efficiencies of the polarizers in the f/96 relay made the FOC a unique instrument, 
the first to take UV polarimetric pictures in space. The FOC remained in operation from 1990 to 2002, when it was replaced by the ACS during Servicing Mission 3B.

\section{The HST/FOC initiative}
Scanning the HST/FOC database, we found that about 15\% of the proposals in the FOC archives lack any exploitation. In addition, all the published observational campaigns do not follow the 
same reduction procedure. Methods vary widely from one group to another, making a rigorous analysis of the whole HST/FOC sample impossible. We thus decided to achieve a meticulous, 
systematically complete and consistent re-analysis of all raw HST/FOC imaging polarimetric AGN observations to enable science deferred or unachieved by many approved programs.

\subsection{A new, generalized reduction pipeline}
In order to standardize the reduction of the HST/FOC data, we followed the steps presented by \citet{Capetti1995a}, \citet{Capetti1995b}, \citet{Nota1996} and \citet{Kishimoto1999} among 
other authors. We dismissed zeroth-order corrections and focused on the most precise operations, always working with the raw data (POL0, POL60, POL120) and Stokes parameters (I,Q,U) to 
estimate and propagate errors. In methodological order, we import the raw data and select the region of interest using a Graham's scan algorithm \citep{Graham1972}. This algorithm finds 
the convex hull of a set of $n$ points in the plane with a complexity $O\left(n \log n \right)$, cropping out undesired values from the data matrix. We then deconvolve the images using the 
Richardson-Lucy deconvolution algorithm \citep{Richardson1972} for recovering the underlying image that has been blurred by the photo-diodes of the detector. After deconvolution, we compute 
errors using a squared zone of the image that is background dominated. This error is later quadratically propagated for any reduction operation on the data. This includes correlated 
error through re-sampling and smoothing. The next steps are data re-sampling, alignment and smoothing thanks to user defined kernels or image combinations. The Stokes parameters are computed 
using the standard \citet{Nota1996}'s procedure, then we derive the polarization degree and angle. Celestial rotation is applied, we compute the proper units for both the total and polarized 
fluxes and display the results before saving everything into FITS files. A very detailed explanation of each of the aforementioned steps will be presented in the first paper of our series 
(Barnouin et al., in prep).

\subsection{Preliminary result: NGC~1068}
In order to test our pipeline, we decided to re-analyze the FOC data of NGC~1068. This is the most archetypal type-2 (edge-on) radio-quiet AGN and thus the best target for benchmarking. 
It possesses the largest database of radio-to-UV polarization measurements \citep{Marin2018} and was even part of the original catalog of Carl Seyfert \citep{Seyfert1943}. NGC~1068 
was observed by the FOC on Feb 28, 1995 (5:33AM), program ID 5144. The dataset was obtained through the F253M UV filter centered around 2530~\AA, together with the polarizing gratings 
POL0, POL60, POL120. The optical relay f/96 was selected to obtain a FoV of 7" $\times$ 7" and a pixel size of 0.014" $\times$ 0.014". It results in a 512 $\times$ 512 pixelated image 
of the source and its environment. Each polarizing grating acquired 3\,500 seconds worth of observation for a total exposure time of 10\,581 seconds (including the CLEAR and F253M filters).

\begin{figure}[ht!]
 \centering
 \includegraphics[width=0.8\textwidth,clip]{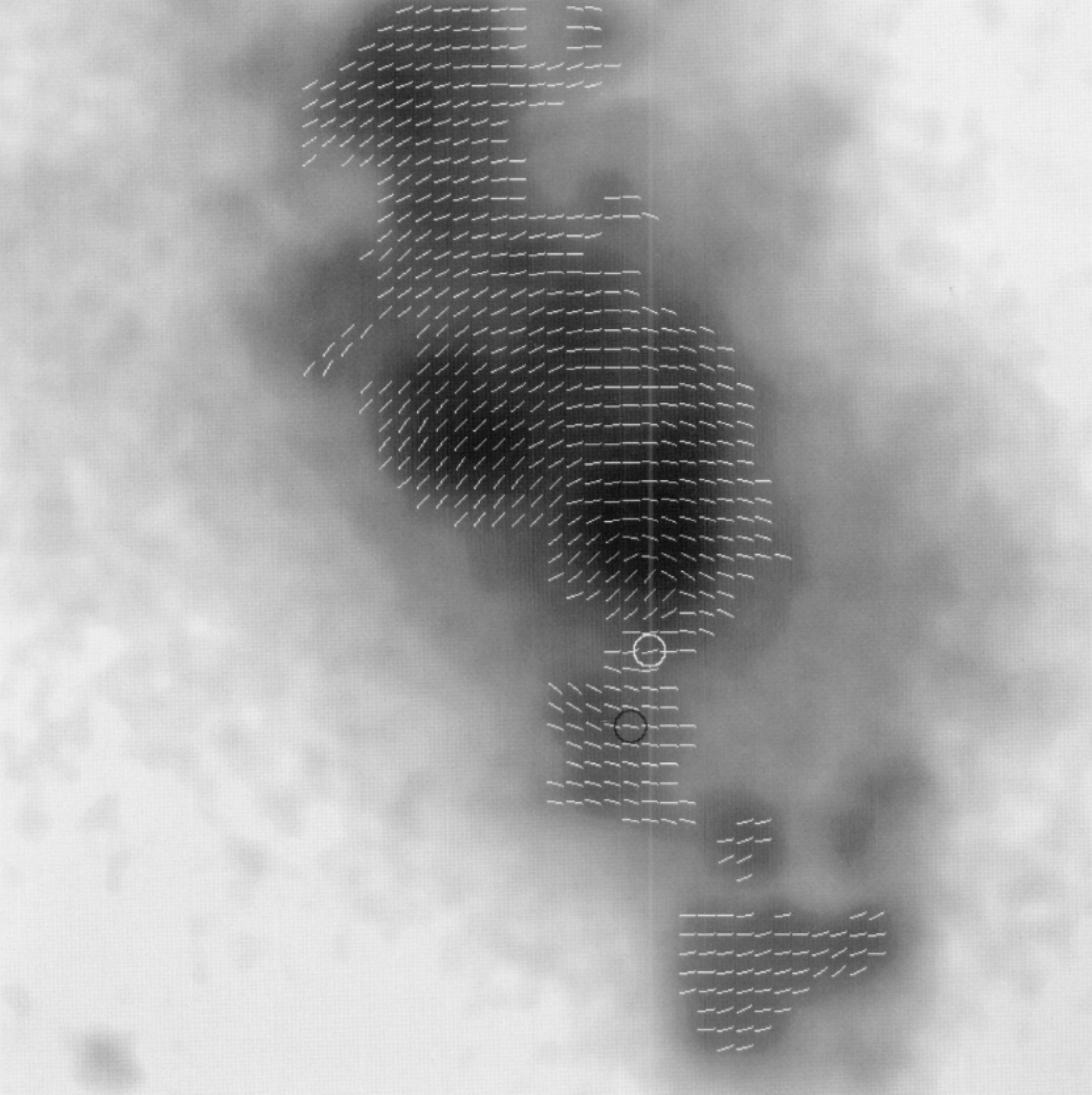}      
  \caption{Total flux image of NGC~1068 obtained by \citet{Capetti1995b}. The polarization information is 
	   superimposed to the image using white vectors. The length of the vectors is fixed and does not 
	   represent the polarization degree value. The polarization position angle can be visualized thanks 
	   to the orientation of the vectors. No indication on the flux strength was provided by the authors.
	   We only know that the darkest regions correspond to the brightest UV fluxes.}
  \label{marin:fig1}
\end{figure}

\begin{figure}[ht!]
 \centering  
 \includegraphics[width=0.8\textwidth,clip]{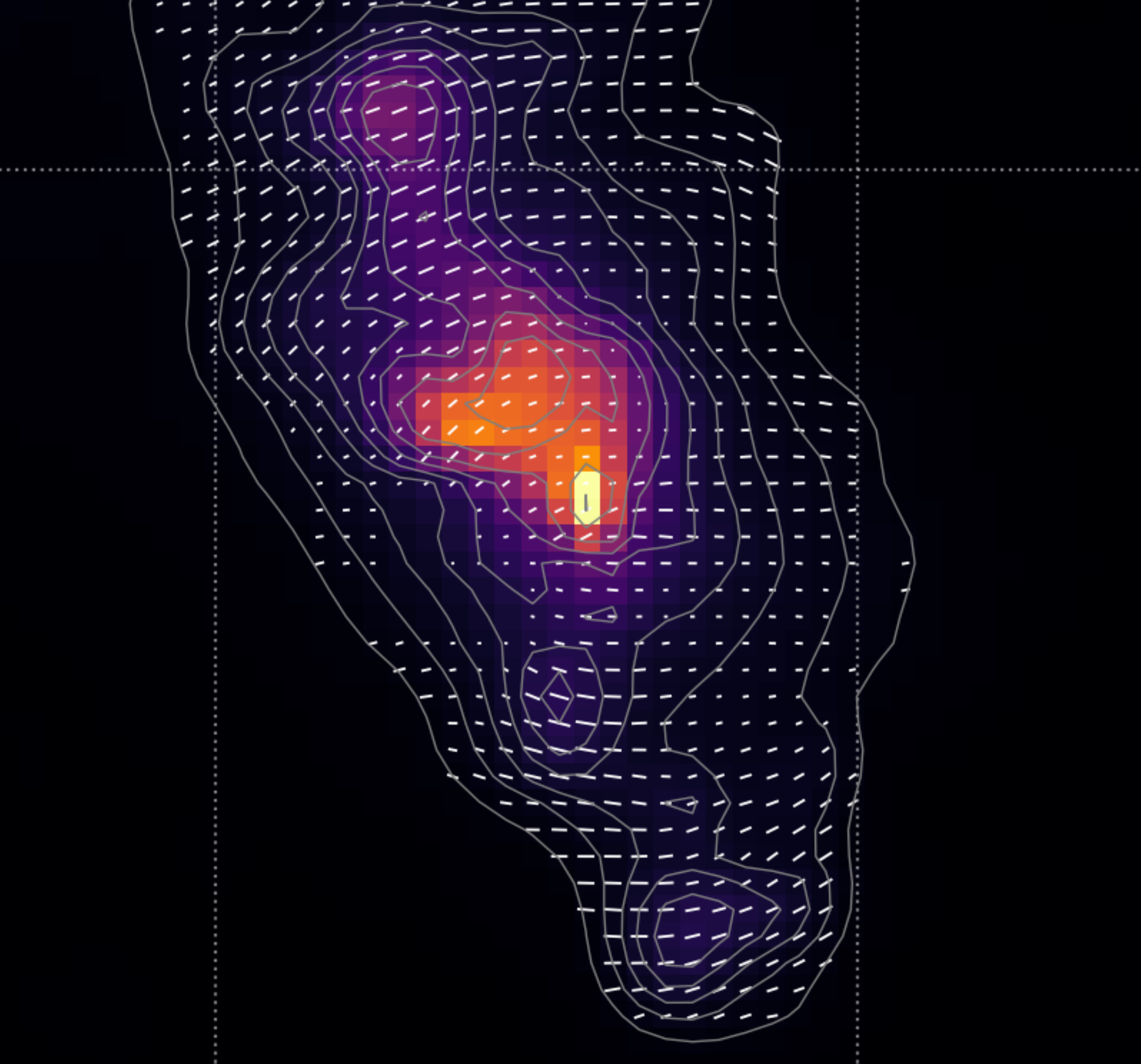}      
  \caption{Results from our re-analysis of NGC~1068. The total flux F$_\lambda$ is linearly color-coded from black 
	   (no flux) to yellow (5.10$^{-16}$~erg.s$^{-1}$.cm$^{-2}$.$\AA^{-1}$). The polarization information is 
	   superimposed to the image using white vectors. The linear polarization degree is proportional 
	   to the vector length while the polarization position angle is indicated by the orientation of the vector 
	   (a vertical vector indicating a polarization angle of 0$^\circ$). North is up, East is left. A spatial bin 
	   corresponds to 0.1". The different contour levels highlight the various signal-to-noise ratios in total 
	   flux in decreasing order from the brightest region to the outskirts of the AGN (433, 399, 366, 332, 298, 
	   265, 231, 197, 164, 130). We intentionally cropped our image to approximatively the same FoV
	   than \citet{Capetti1995b}'s, see Fig.\ref{marin:fig1}.}
  \label{marin:fig2}
\end{figure}

Fig.~\ref{marin:fig1} shows the total flux image of NGC~1068 obtained by \citet{Capetti1995b} with the Feb 28, 1995's observation. Fig.~\ref{marin:fig2} presents the exact same observation
processed through our standardized pipeline. Despite being larger and containing more polarimetric information, we intentionally cropped our figure to approximatively the same FoV 
(3.3 $\times$ 2.9 arcsec) than the figure from Capetti to make comparisons easier. Our figure is also polychromic. Apart from those visual details, the global flux mapping and polarization 
vector pattern are extremely similar between the two figures. It is reassuring, telling us that our method works smoothly. But striking differences also appear. First, we extracted much 
more information from the raw data in terms of polarization. For the same cut-off in signal-to-noise ratio in polarization ($>$20), we detect much more polarized pixels. Those pixels, mostly 
along the polar direction of the AGN symmetry axis, better highlight the half-opening angle of the outflows. This could be used to constrain the half-opening angle of the circumnuclear, 
dusty region if the winds fill the whole solid angle. The vectors also define the centro-symmetric shape of the polarization pattern with a greater precision than before, validating the 
approach taken by authors such as \citet{Gratadour2015} to reveal obscured regions by contrast. Finally, small details vary from Fig.~\ref{marin:fig1} to Fig.~\ref{marin:fig2}, such as the 
position of the maximum polarization degree pixels and the presence of vectors misaligned with respect to the centro-symmetric pattern at the outskirts of the polar winds. It is too early to determine 
if this is due to scattering on a dust lane \citep{Stalevski2017,Vollmer2018}, background contamination by the host \citet{Marin2018b} or the presence of parsec-scale magnetic fields \citet{Lopez-Rodriguez2020}.

\section{Conclusions}
We have developed a new, standardized, consistent reduction pipeline to explore in great details the archives of the HST/FOC. Because 15\% of the observations were never reduced, and because 
the rest has not been explored in a consistent way, we aim at producing a complete and generalized catalog with high-resolution images for the community. We expect new discoveries and refined
conclusions on the geometry of the scattering regions in AGNs, the detection and location of hidden nuclei, the composition of the polar outflows (electrons, dust or a mixture of both) and 
novel constrains on the evolution of galaxies.

\begin{acknowledgements}
The authors would like to thank Dr. Makoto Kishimoto for his continuous help in establishing a robust and standardized reduction pipeline.
\end{acknowledgements}

\bibliographystyle{aa}  
\bibliography{marin_S19}

\begin{thebibliography}{16}
\expandafter\ifx\csname natexlab\endcsname\relax\def\natexlab#1{#1}\fi

\bibitem[{{Capetti} {et~al.}(1995{\natexlab{a}}){Capetti}, {Axon}, {Macchetto},
  {Sparks}, \& {Boksenberg}}]{Capetti1995a}
{Capetti}, A., {Axon}, D.~J., {Macchetto}, F., {Sparks}, W.~B., \&
  {Boksenberg}, A. 1995{\natexlab{a}}, \apj, 446, 155

\bibitem[{{Capetti} {et~al.}(1995{\natexlab{b}}){Capetti}, {Macchetto}, {Axon},
  {Sparks}, \& {Boksenberg}}]{Capetti1995b}
{Capetti}, A., {Macchetto}, F., {Axon}, D.~J., {Sparks}, W.~B., \&
  {Boksenberg}, A. 1995{\natexlab{b}}, \apjl, 452, L87

\bibitem[{{Code} \& {Welch}(1979)}]{Code1979}
{Code}, A.~D. \& {Welch}, G.~A. 1979, \apj, 228, 95

\bibitem[{{Davidsen}(1993)}]{Davidsen1993}
{Davidsen}, A.~F. 1993, Science, 259, 327

\bibitem[{{Graham}(1972)}]{Graham1972}
{Graham}, R. 1972, Information Processing Letters, 1, 132

\bibitem[{{Gratadour} {et~al.}(2015){Gratadour}, {Rouan}, {Grosset},
  {Boccaletti}, \& {Cl{\'e}net}}]{Gratadour2015}
{Gratadour}, D., {Rouan}, D., {Grosset}, L., {Boccaletti}, A., \& {Cl{\'e}net},
  Y. 2015, \aap, 581, L8

\bibitem[{{Kishimoto}(1999)}]{Kishimoto1999}
{Kishimoto}, M. 1999, \apj, 518, 676

\bibitem[{{Linsky}(2018)}]{Linsky2018}
{Linsky}, J.~L. 2018, \apss, 363, 101

\bibitem[{{Lopez-Rodriguez} {et~al.}(2020){Lopez-Rodriguez}, {Dowell}, {Jones},
  {Harper}, {Berthoud}, {Chuss}, {Dale}, {Guerra}, {Hamilton}, {Looney},
  {Michail}, {Nikutta}, {Novak}, {Santos}, {Sheth}, {Siah}, {Staguhn},
  {Stephens}, {Tassis}, {Trinh}, {Ward-Thompson}, {Werner}, {Wollack},
  {Zweibel}, \& {HAWC+Science Team}}]{Lopez-Rodriguez2020}
{Lopez-Rodriguez}, E., {Dowell}, C.~D., {Jones}, T.~J., {et~al.} 2020, \apj,
  888, 66

\bibitem[{{Marin}(2018{\natexlab{a}})}]{Marin2018}
{Marin}, F. 2018{\natexlab{a}}, \mnras, 479, 3142

\bibitem[{{Marin}(2018{\natexlab{b}})}]{Marin2018b}
{Marin}, F. 2018{\natexlab{b}}, \aap, 615, A171

\bibitem[{{Nota} \& {et al.}(1996)}]{Nota1996}
{Nota}, A. \& {et al.} 1996, {Faint Object Camera Instrument Handbook v.7}

\bibitem[{{Richardson}(1972)}]{Richardson1972}
{Richardson}, W.~H. 1972, J. Opt. Soc. Am., 62, 55

\bibitem[{{Seyfert}(1943)}]{Seyfert1943}
{Seyfert}, C.~K. 1943, \apj, 97, 28

\bibitem[{{Stalevski} {et~al.}(2017){Stalevski}, {Asmus}, \&
  {Tristram}}]{Stalevski2017}
{Stalevski}, M., {Asmus}, D., \& {Tristram}, K. R.~W. 2017, \mnras, 472, 3854

\bibitem[{{Vollmer} {et~al.}(2018){Vollmer}, {Schartmann}, {Burtscher},
  {Marin}, {H{\"o}nig}, {Davies}, \& {Goosmann}}]{Vollmer2018}
{Vollmer}, B., {Schartmann}, M., {Burtscher}, L., {et~al.} 2018, \aap, 615,
  A164

\end{thebibliography}

\end{document}